\documentclass[11pt,a4paper]{article}
\usepackage{jheppub}
\usepackage{amsmath}
\usepackage{amsfonts}
\usepackage{amssymb}
\usepackage{graphicx}
\newcommand{\tb}{\tilde{b}}
\newcommand{\tc}{\tilde{c}}
\preprint{DIAS-STP-18-02}

\title{Triple Point of a Scalar Field Theory on a Fuzzy Sphere}

\author[]{Samuel Kov\'a\v{c}ik,}
\author[]{Denjoe O'Connor,}
\affiliation[]{School of Theoretical Physics,\\ 
       Dublin Institute for Advanced Studies, \\
       10 Burlington Road, 
       Dublin 4, Ireland.}

\emailAdd{skovacik@stp.dias.ie}
\emailAdd{denjoe@stp.dias.ie}

\abstract{The model of a scalar field with quartic self-interaction on
  the fuzzy sphere has three known phases: a uniformly ordered phase,
  a disordered phase and a non-uniformly ordered phase, the last of
  which has no classical counterpart. These three phases are expected to
  meet at a triple point. By studying the infinite matrix size
  limit, we locate the position of this triple point to within a small
  triangle in terms of the parameters of the model.  We find the
  triple point is closer to the coordinate origin of the phase diagram
  than previous estimates but broadly consistent with recent
  analytic predictions.

\vspace*{.3cm}

\vspace*{.3cm}

}

\begin{document}

\maketitle

\section{Introduction}
Noncommutative coordinates appear in various areas of
physics. Sometimes they emerge as an effective description of a
physical system \cite{hall, Fujii:2005kg} or as a fundamental
structure of the underlying space \cite{Grosse:1996mz,
  Ydri:2001pv,Grosse:1998gn,DiFrancesco:1993cyw}. Most notably, many
of the candidate theories for quantum gravity predict such structure
when approaching the Planck scale \cite{Doplicher:1994tu,Seiberg:1999vs,Witten:1986}.

Though noncommutativity was initially thought to provide a
regularisation of field theories, it was quickly realised that this is
not the case. In its fuzzy incarnation \cite{Madore:1991bw}, where the number of
degrees of freedom is finite, it can still be an effective
regularisation scheme; especially in bosonic theories, where the
underlying symmetries, for example, the rotational symmetry in the case
of the fuzzy sphere, are preserved. Such theories are
nonlocal and typically suffer from the infamous UV/IR mixing
problem \cite{Chu:2001xi}. As a result, phase diagrams of these
theories differ from their commutative counterparts (even after taking
what one would expect to be the commutative limit).

Noncommutative
(or fuzzy) spaces and field theories on them have been studied
extensively both analytically
\cite{Dolan:2001gn,Chu:2001xi,Tekel:2017nzf,Polychronakos:2013nca,Tekel:2016jbp,Tekel:2015zga,Tekel:2015uza,Grosse:1995jt,Iso:2001mg,OConnor:2007pvc,OConnor:2007ibg,Rea:2015wta,CarowWatamura:1998jn,Madore:1991bw}
and numerically
\cite{GarciaFlores:2009hf,Ydri:2014rea,Ydri:2015vba,Panero:2006bx,Ydri:2015zba,Vachovski,Sabella-Garnier:2017svs}.
A good example is the scalar field theory with a quartic
potential on a fuzzy sphere, i.e. the fuzzy $\Phi^4$ model. The
phase diagram is controlled by the triple point, from which the phase
boundaries radiate outwards and divide the plane into three distinct phase
regions. The goal of this paper is to locate the triple point of the
fuzzy $\Phi^4$ model as precisely as possible.

Earlier studies \cite{GarciaFlores:2009hf,Ydri:2014rea} used rather
small matrix sizes and linear extrapolation from relatively far from
the triple point to infer its location. We provide a reasonably
precise estimate of its position in the $N\rightarrow \infty$ limit
and find good evidence that the three known phases of the model do in
fact meet at a well defined triple point.  We find it to be located
much closer to the origin than previous studies estimated, but in
reasonably good agreement with results using an analytic approximation
initiated in \cite{OConnor:2007ibg} and recently further developed in
\cite{Tekel:2017nzf}; for other studies in this direction see
also \cite{Polychronakos:2013nca,Rea:2015wta}.

The paper is structured as follows. Section \ref{TheModel} describes the
$\Phi^4$ model on a fuzzy sphere. Section \ref{Methods} describes our
approach and basic results. The final section is devoted to our conclusions.

\section{The fuzzy $\phi^4$ model}
\label{TheModel}
A real scalar field theory with $\phi^4$ potential on a (commutative)
sphere of radius $R$ is described by the action
\begin{equation} \label{actionC}
S[\phi] = \int \limits_{\textbf{S}^2}d\Omega \left[ \frac{1}{2}\phi \Delta \phi  + \frac{r R^2}{2} \phi^2 + \frac{g R^2}{4!} \phi^4 \right],
\end{equation}
where $\Delta=-\nabla^2$ is the Laplacian (with positive spectrum)
on the sphere, $r$ is the
mass parameter (which can be negative), $g$ is the
coupling constant and $d\Omega$ is the volume form on the unit two-sphere.
When $r>0$ the potential is minimised by a
configuration with $\langle \phi \rangle =0$. For sufficiently small
(and negative) $r$, the potential has two minima
$\langle \phi \rangle= \pm \phi_0$; in the infinite volume limit
the system spontaneously breaks the ${\mathbb Z}_2$
symmetry and oscillates around one of them. When statistical
fluctuations are taken into account the model is in the universality
class of the two dimensional Ising model and the transition line is
described by a conformal field theory. Lattice simulations of the
two-dimensional commutative model ($\phi^4_2$) on the plane were performed
in \cite{Loinaz:1997az,Bosetti:2015lsa}, where the transition line
in the $(r,g)$ plane was measured. A recent study of the
two-dimensional commutative model was performed using a finite element
approximation to the two-sphere in \cite{Brower:2016moq} and they
find good agreement with results expected from conformal field theory
and renormalisation group studies.

We are interested in the non-commutative version of this scalar field
defined on a fuzzy sphere specified by the action
\begin{equation} \label{action}
S_N[\Phi]=\mbox{Tr } \left[ a\ \Phi^\dagger \mathcal{L}^2 \Phi + b \ \Phi^2 +  c \ \Phi^4 \right],
\end{equation}
where $\Phi$ is a Hermitian $N\times N$ matrix, $\mathcal{L}^2 \Phi = [L_i, [ L_i , \Phi ]]$ with $L_i, i=1, 2, 3$ being the usual $SU(2)$ generators in the irreducible representation of dimension $N$. This action can be seen as a regularisation of the commutative model by defining the matrix
$\Phi$ associated with a field configuration $\phi$ as
\begin{equation}
\Phi=\int  \limits_{\textbf{S}^2} d\Omega \ \rho(\hat{n})\phi(\hat{n}),\quad \hbox{where}\quad\rho(\hat{n})=\sum_{l,m=0}^{N-1}\hat{Y}_{lm}Y^\dag_{lm}(\hat{n}),
\label{maptofuzzy}
\end{equation}
$Y_{lm}(\hat{n})$ are the standard spherical harmonics and
$\hat{Y}_{lm}$ are the polatisation tensors; details of the map
$\rho(\hat{n})$ can be found in \cite{Dolan:2006tx}. Its effect can be understood by expanding the field $\phi$ in spherical
harmonics. The map (\ref{maptofuzzy}) replaces spherical harmonics
with polarisation tensors up to the maximum angular momentum allowed
by the matrix size, i.e. $L=N-1$, higher modes being cut off. With
this identification one can establish that
\begin{equation}
\lim_{N\rightarrow\infty} \left\vert S[\phi]-S_N[\Phi]\right\vert\rightarrow 0
\end{equation}
provided we identify $a = \frac{2\pi }{N}, b=\frac{2 \pi r R^2}{N}$ and $c= \frac{\pi g R^2}{6N}$.

From the above construction, we see that a field on a fuzzy sphere (or
any other similarly constructed fuzzy space) has only a finite number
of degrees of freedom, but contrary to a lattice discretisation, the
underlying rotational symmetry is preserved.

We are interested in the statistical mechanics of this model where the mean
value of an observable is defined as 
\begin{equation} \label{mean}
\langle \mathcal{O} (\Phi)\rangle = \frac{\int d [\Phi] \mathcal{O}(\Phi) e^{-S_N[\Phi]}}{\int d [\Phi] e^{-S_N[\Phi]}},
\end{equation}
and we integrate over all possible (Hermitian) matrix configurations.
Without loss of generality, we can rescale the matrix $\Phi$ to make $a=1$ and
use the scale-free parameters $\tilde{b} = b/(a N^{\frac{3}{2}}),\ \tilde{c}=c/(a^2 N^2)$ as in \cite{GarciaFlores:2009hf}.

The finiteness of the number of degrees of freedom is not the only novel
feature of the matrix model compared to its commutative
counterpart. If there was no gradient term in \eqref{action}, the
action could be expressed in terms of the eigenvalues,
$\lambda_i$, of $\Phi$. The integration would become $\int d [\Phi] \rightarrow
\int \Delta^2(\lambda) d [\lambda] d [\theta]$, where
$\Delta(\lambda)$ is the Vandermonde determinant.

As long as the observable in question depends only on
$\lambda_i$, the integral $\int d [\theta$] gives the volume of
the corresponding flag manifold and cancels in \eqref{mean}. The Vandermonde
determinant lifted to the exponential acts as an additional logarithmic
repulsion between the eigenvalues. In the large $N$ limit
the pure potential model has a third order phase transition,
at $b_c=-2\sqrt{N c}$,  to
a phase where the eigenvalues are distributed
between both minima of the potential and do not jump the
barrier. This new phase is closely related to the additional phase present in
the model \eqref{action}.

The overall form of the phase diagram for the $\Phi^4$ model on a fuzzy sphere is the following: when $\tilde{c}$ is sufficiently large (and fixed), there are two critical values for $\tb$. For $\tb > \tb_{c_1}$, the system oscillates
around the trivial vacuum,
$\langle\lambda_i \rangle=0$. The eigenvalue distribution in this phase
is a distorted Wigner semicircle -- this is called the disordered phase.  On
the other hand, when $\tb < \tb_{c_2}$, all of the eigenvalues
oscillate around the minimum of one of the pair
of (deep) potential wells related by ${\mathbb Z}_2$ symmetry,
$\langle\lambda_i \rangle=\pm\lambda_0$. Due
to the repulsion between the eigenvalues discussed above,
when $\tb_{c_2}<\tb<\tb_{c_1}$ and $\tc$ is sufficiently large, there is an
additional matrix phase, which lies between the disordered and
the uniform phase. It has no classical counterpart in the commutative model and is
characterised by some eigenvalues in one well and rest of them in the other,
$\langle\lambda_i \rangle={\pm}_i \lambda_0$. This phase is referred to
as non-uniformly ordered. 

As the coupling $\tc$ is lowered, the two transitions at $\tb_{c_1}$ and $\tb_{c_2}$
approach and eventually meet
at the triple point $(\tc_T, \tb_T)$. For $\tc < \tc_T$, the
non-uniformly ordered phase is absent. It is the location of this triple
point we seek.

\section{Methods and results}
\label{Methods}

We performed hybrid Monte Carlo (HMC) \cite{Duane:1987de} simulations
of the model with the probability weight defined by the action
\eqref{action}. In contrast to the ordinary Monte Carlo method, the
hybrid one includes molecular dynamics that shortens the
auto-correlation length and the required computational time.

The idea of the HMC method (see e.g. 
  \cite{HMC}) is that instead of just
  updating the configuration with a random impulse, one first
  introduces fictional momentum degrees of freedom and treats the
  original action as the potential for these momenta in a new
  Hamiltonian.  The integral over the new momenta is trivial as they
  are Gaussian degrees of freedom. One then finds a proposed update
  for $\Phi$ as a solution to the discretised Hamilton equations,
  where the leapfrog interval and number of steps are new
  parameters of the simulation.  One then imposes the Metropolis check
  on the proposed configuration. The resulting configurations are then
  much more decorrelated and the system thermalises more rapidly. 

In a comparison of the Metropolis and HMC methods we
  ran test simulations generating $10^5$ configurations with $N=15$,
  $\tb = -0.1$, $\tc =0.5$. To facilitate a comparison the Metropolis
  algorithm was implemented with a complete matrix updated by
  proposing a new configuration
  $\Phi'=\Phi+\epsilon_{\mbox{\tiny{M}}} P$ with $P$ choosen from a
  Gaussian distribution.  We tuned the
  $\epsilon_{\mbox{\tiny{M}}}$ and the HMC step-length 
  $\epsilon_{\mbox{\tiny{HMC}}}$ to achieve the same acceptance rate
  ($ 90\%$ for this example only) for the Metropolis
  and the HMC algorithms. While the former required considerably
  shorter step-lenght,
  $\epsilon_{\mbox{\tiny{M}}}\simeq\frac{\epsilon_{\mbox{\tiny{HMC}}}}{60}$,
  it also produced data with
  considerably longer auto-correlation length,
  $\lambda_{\mbox{\tiny{M}}}\simeq 14\lambda_{\mbox{\tiny{HMC}}}$.

In actual simulations we tried to keep the acceptance
  rate between 60\% -- 90\% which differs slightly from the recommended
  value (we wanted to use the same value for large portions of the
  phase diagram to make the simulations as automated as possible) and auto-correlation lengths of the order
  of ten configurations or less. We
  typically ran $\sim10^3$ values of $\tilde{b}$ and $\tilde{c}$
  to construct a phase diagram
  with fixed $N$, each of them generating $10^5-10^6$ configurations
  (dropping the first $\sim10^4$ steps of thermalisation). 

The HMC algorithm allowed us the work with larger
matrices, up to $N=127$, which we have used to extrapolate to
$N\rightarrow \infty$. We have measured the value of the action
for each of the Monte
Carlo steps, from which the specific heat
$C_v = \left( \langle S^2 \rangle - \langle S \rangle^2 \right)/N^2$
and the Jack-knife error estimate were obtained.

Since we are dealing with a finite size system the phase transitions
are smeared and manifest themselves as (finite) peaks in the specific
heat.

\begin{figure}[htb]

\centering

    \includegraphics[width=1\textwidth]{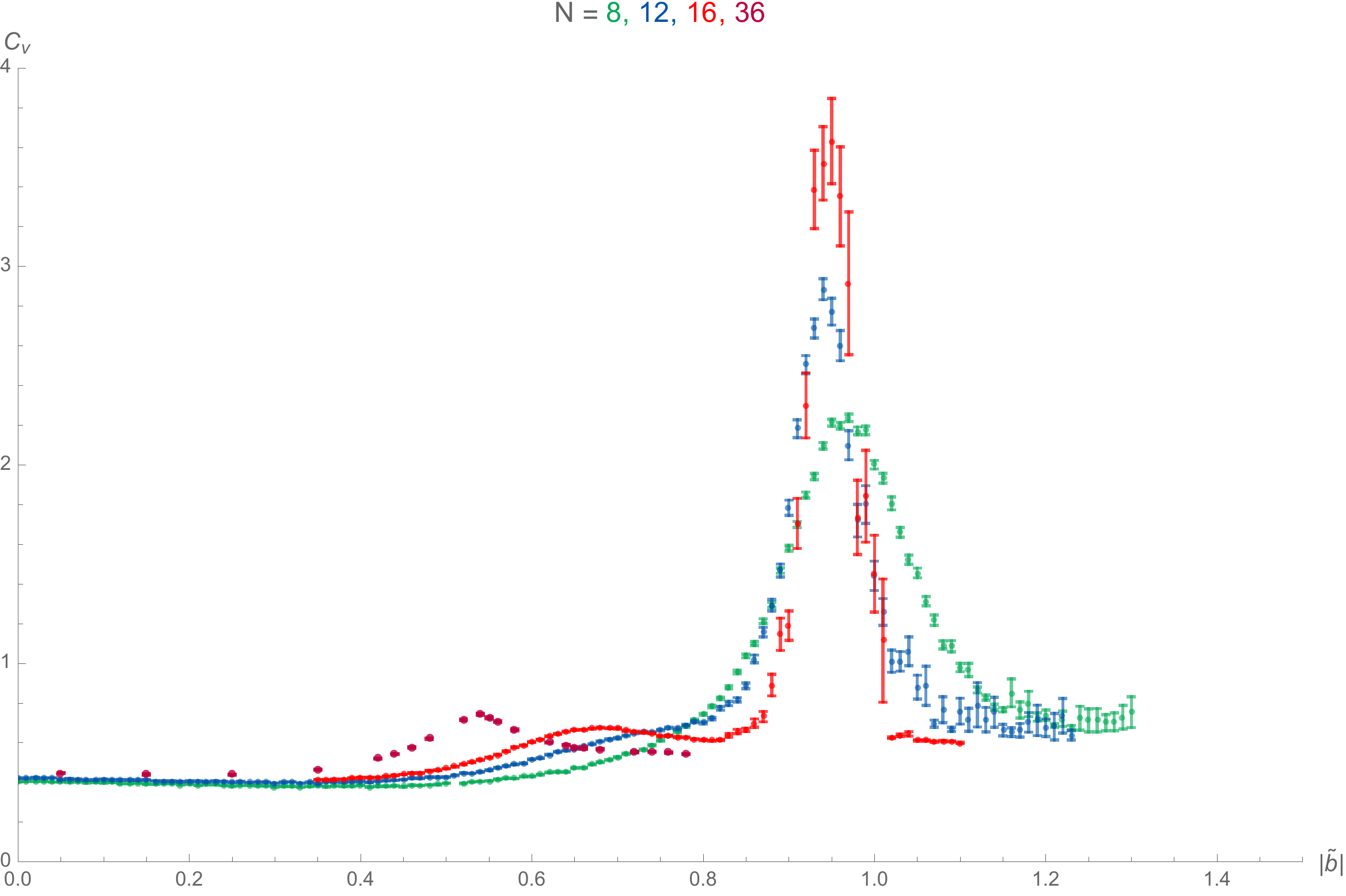} \\

    \caption{The specific heat for small $N$ and $\tc=0.20$. In the
      $N=16$ plot only the main peak corresponding
      to the non-uniformly to uniformly ordered phase boundary is
      visible. The plots for $N=12$ and $N=16$ show the emergence of
      the secondary peak corresponding to the disordered to
      non-uniformly ordered phase boundary. For $N=36$ only the latter peak is
      visible.}
\label{CvSmallN}
\end{figure}

The strategy was to fix the value of $N$ and $\tc$ and to make
simulations with multiple, slightly separated, values of $\tb$. This
allowed us to plot $C_v(\tb)$ and identify the phase transition values
of $\tb$, see Figure \ref{CvSmallN}. The number of data-points
(simulations) for the analysed
values of $N$ was of the order of thousands, which was sufficient to
construct the phase diagram with the required precision.

\begin{figure}[!tbp]
    \includegraphics[width=\textwidth]{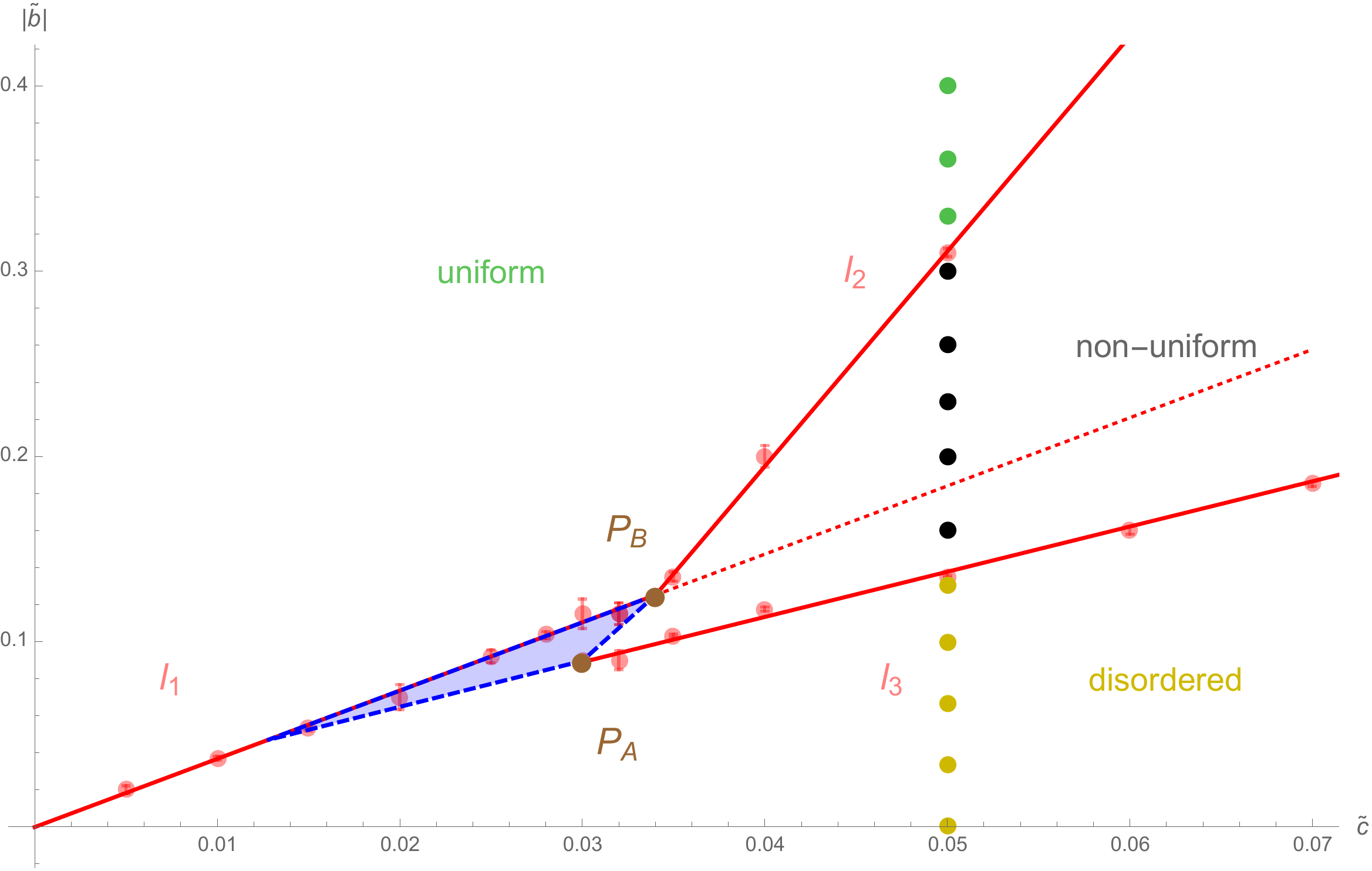}
    \caption{The phase diagram for $N=64$. Red lines are the fits of
      measured phase transitions, which are denoted by points with
      barely visible error bars. $P_B$ denotes the point where $l_1$
      bends into $l_2$, $P_A$ denotes the point where $l_3$
      emerges. The position of the triple point can be estimated to
      either lay in the blue triangle or, as a cruder estimate, just
      between these two points. The phase transitions were identified
      from the behaviour of the specific heat $C_v$, but for $\tc =0.05$
      we have also checked the eigenvalue distribution directly
      to verify it (the colour points correspond to their respective
      phases). The lines are fitted as
      $l_1: |\tb| = -0.000(2) + 3.68(9) \tc$,
      $l_2: |\tb| = -0.27(2) + 11.6(4) \tc$,
      $l_3: |\tb|= 0.015(2) + 2.44(4) \tc$. }
\label{N64PhaseDiag}
\end{figure}

The overall structure of the phase diagram is the same for all
values of $N$. There is one coexistence curve starting from the origin
of the phase diagram separating the disordered and the uniformly
ordered phases, denoted $l_1$. It bends to a new curve $l_2$ at a
point labeled $P_B$.  Close to $P_B$ a new coexistence line, $l_3$,
appears at a point we denote $P_A$. The non-uniformly ordered phase exists
between $l_2$ and $l_3$, the uniformly ordered phase exists above the
line $l_2$ and the disordered one below $l_2$, see Figure \ref{N64PhaseDiag}.
All of the coexistence curves, $l_1,l_2$ and $l_3$, are
well approximated by straight lines, at least in the vicinity of
the triple point.

The line $l_2$ in Figure \ref{N64PhaseDiag} is more correctly
described as the set of values where the potential barrier separating
the two phases has become too large for the fluctuations of the system
to successfully jump between the phases.  The specific heat in this
region would be extremely large and the system can be modeled by two
Gaussian distributions whose means differ but have similar standard
deviations. Since the value of $\tb$ at which this occurs approaches
the triple point with increasing $N$, this line provides a useful
marker of the upper boundary of the triangles we use to locate the
triple point. It should not be taken as the true coexistence
curve. Rather the coexistence curve is a continuation of the line
$l_1$ and we find that the two coexistence curves in the vicinity of
the triple point fall on the one, surprisingly straight, line through
the origin. 

One can see the peaks in $C_v$ for different $N$ and
fixed $\tc=0.20$ in Figure \ref{CvSmallN}. For $N=16$ both peaks are
present and one can also observe that the location of the larger peak,
corresponding to the disorder-nonuniform order coexistence line, is
largely insensitive to $N$ but grows rapidly with increasing $N$.
Our results agree with those of the earlier studies in the region
of the phase diagram accessed in those studies.

As analysed in \cite{Tekel:2017nzf}, both uniform and non-uniform
solutions coexist and the dominant phase is determined by the one of
lower free energy. In the vicinity of the coexistence line
the two states are separated by a high barrier, and the simulation
can spend a large number of Monte Carlo steps oscillating around
a wrong vacuum state; this makes the line $l_2$ hard to measure
precisely--for larger $N$ it is especially
difficult to maintain ergodicity in the simulations.

 \begin{figure}[!tbp]
   \includegraphics[width=\textwidth]{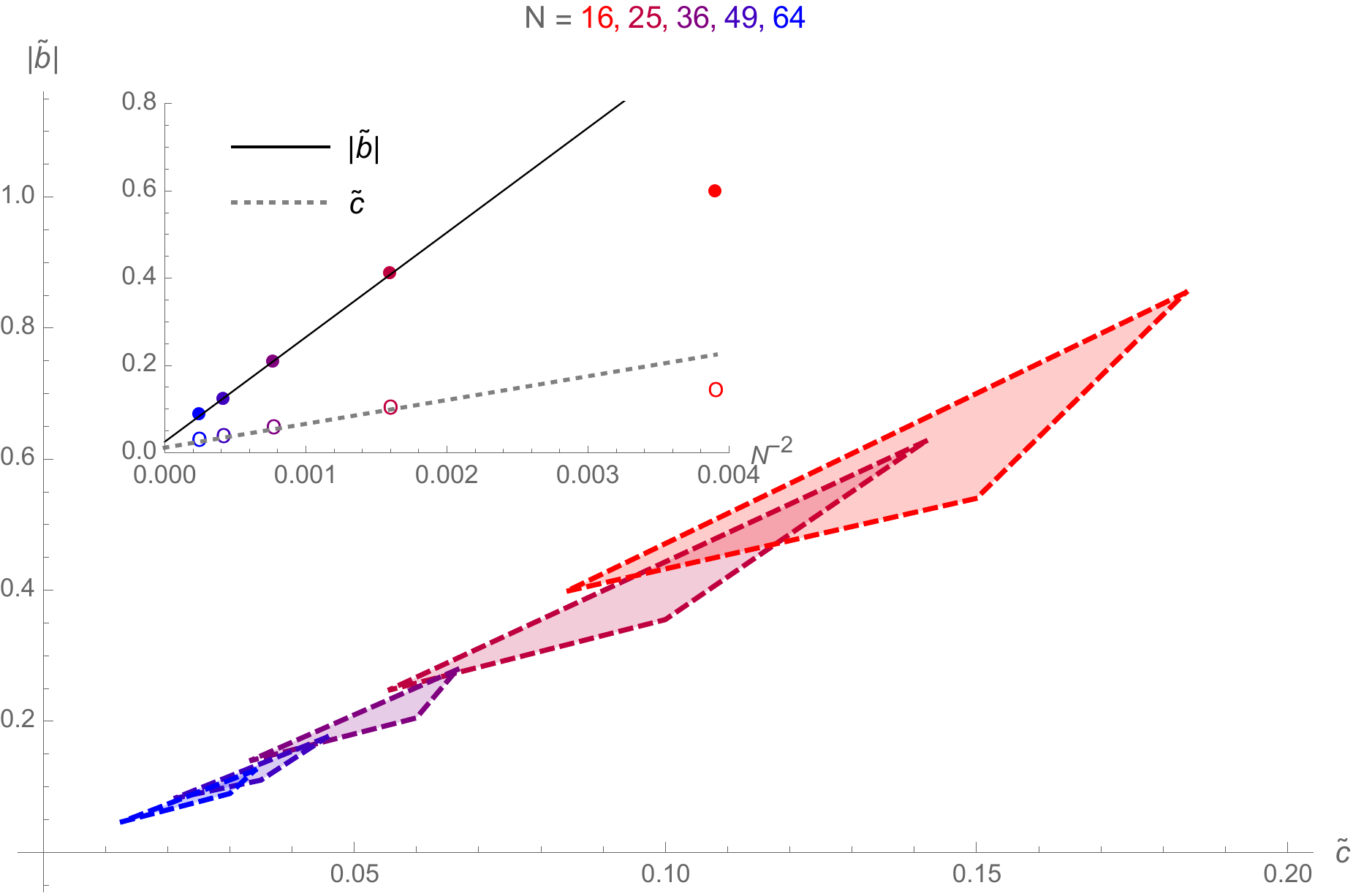}
    \caption{The progression of triangles $t_N$ used to estimate the
      position of the triple point for various values of $N$. The
      results for $N=16$ was not used in the extrapolation as this
      value of $N$ seems to be too small and this result differs from
      the clear trend set by the rest of the values, which are nearly
      perfectly fitted by a linear function of $N^{-2}$, see the small
      statistical errors in Table \ref{tab:results}.}
\label{TrianglesConvergeToTriplePoint}
\end{figure}

The lines $l_1, l_2, l_3$ do not intersect exactly, which is
partially caused by the finiteness of $N$ and partially by the statistical
errors of the measured data. Therefore, instead of meeting, they form a
triangle $t_N = \{P_1, P_2, P_3\}$, where by definition $P_2 \approx
P_B$ and from our experience $P_3 \approx P_A$.

It is impossible to localize the triple point exactly with a finite
value of $N$, but it can be expected to be contained in the triangle
$t_N$. We have observed that the area of $t_N$ declines as $N$ grows
and the center of mass of its upper-half converges to a certain value -- which we take to
be the position of the triple point in the $N\rightarrow \infty$
limit. The systematic error of this estimate is the size (the square
root of the area) of the triangle, the statistical error is that of
the fitting coefficients (the coordinates of the triple point are well
fitted as linear functions of $N^{-2}$).

A more straightforward and alternative approach would be to take the
estimate of the triple point to be the middle-point between
$P_A$ and $P_B$, the error being a half of their distance.

Another option is first to extrapolate the phase transition points for
fixed $\tc$ (in the $\tb$ direction) and to construct the phase
transition lines only afterward. The advantage of this method is that
it allows us to measure the linear coefficients of the curves in the
vicinity of the tripe point. The disadvantage is that different points
converge with increasing $N$ at different rates. For example, the
disordered to non-uniformly ordered transition close to the triple
point requires very large matrix sizes, for our closest point even
$N=127$ seemed insufficient.

\begin{figure}[htb]
\centering
\label{extrapolated}

    \includegraphics[width=1\textwidth]{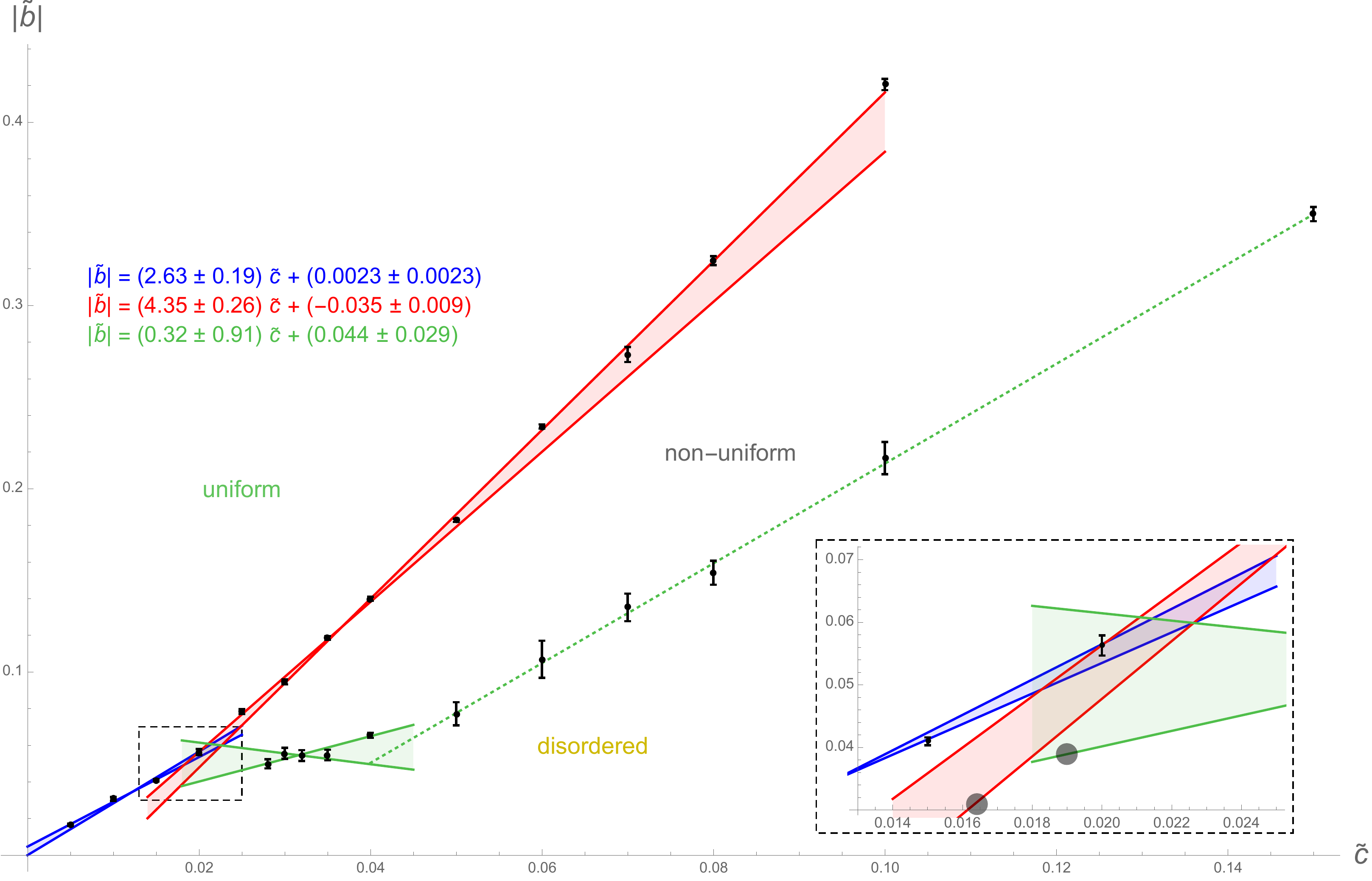}

  \caption{Phase diagram obtained by first extrapolating the phase transition points $\tb(\tc)$ as $N \rightarrow \infty$ and then by fitting the lines close to the triple point. There are two or three curves for each of the transition lines because their slopes change as the triple point is approached. The coloured region is bounded by two lines -- one obtained by fitting only the nearest well-resolved points, the other using also more distant ones; the correct phase transition lines should be bounded by those lines. Center of the area where these regions overlap is quoted as the third triple point estimate in Table \ref{estimates}. The values of used matrix sizes differ for each of the points, but were in the range $8 \le N \le 127$. There is an interesting point where three of the boundary curves cross: $\tc = 0.0226, |\tb| = 0.0599$. The grey points in the zoomed plot mark the estimates obtained using the two other methods.}
\label{EigenvalueaAndInvCv}
\end{figure}

Results of these three methods are gathered in Table
\ref{estimates}. Coefficients of the curves in the $N \rightarrow
\infty$ limit can be found in Figure \ref{extrapolated}.

\begin{table}
\caption {Estimation of the triple point.} \label{tab:results} 
\label{estimates}
\begin{center}
\begin{tabular}{|c|l|l|} 
\hline 
• & \hspace{2cm}$\tc$ &\hspace{2cm}$|\tb|$ \\ 
\hline 
Triangle &  $0.016  \pm 0.009_{\mbox{sys}} \pm 0.001_{\mbox{st}}$ & $- 0.031 \pm 0.009_{\mbox{sys}} \pm 0.009_{\mbox{st}}$ \\ 
\hline 
Middle-point & $0.019  \pm 0.001_{\mbox{sys}}\pm 0.002_{\mbox{st}}$ & $ - 0.039\pm 0.015_{\mbox{sys}} \pm 0.005_{\mbox{st}} $ \\ 
\hline 
Extrapolation & $0.021 \pm 0.002_{\mbox{sys}}$ & $- 0.057 \pm 0.006_{\mbox{sys}}$ \\ 
\hline 
\end{tabular} 
\end{center}
\end{table}

The previous study reported the values of $\tc = 0.52, \ \tb = -2.3$
(see \cite{GarciaFlores:2009hf}) and $\tc = 0.58, \ \tb = -2.49$ (see
\cite{Ydri:2014rea}), but no data was gathered in the immediate vicinity of the transition due to the smallness of the matrices studied, rather the
triple point was inferred from a linear extrapolation of the three coexistence curves obtained far from the triple point. The maximal matrix size used in those studies was $N=10$,
which is, as we have learned, not free of small $N$ effects (when
plotting the estimates of $\tc, |\tb|$ versus $N^{-2}$, even the
point $N=16$ deviates from the very clear trend set by $N=25, 36,49,64$;
see Figure \ref{TrianglesConvergeToTriplePoint}). Also, we were able to measure the phase transition
closer to the triple point, not having to rely on extrapolations from
distant regions.

The new estimates for the triple point position is in reasonable accord
with a recent analytic study \cite{Tekel:2017nzf} which provides the
values of $\tc_T =0.0196, \ \tb_T = -0.35$. This value of $\tc$ agrees very well with our estimate but the discrepancy in $\tb$
is significant and lies well outside our errors.  It is probably caused by the approximations
used in \cite{Tekel:2017nzf} and further effort on the analytic
side is expected to improve the agreement.\footnote{Private communication with the author.}
 
 \begin{figure}[htb]
\centering

  \begin{tabular}{@{}c@{}}

    \includegraphics[width=1\textwidth]{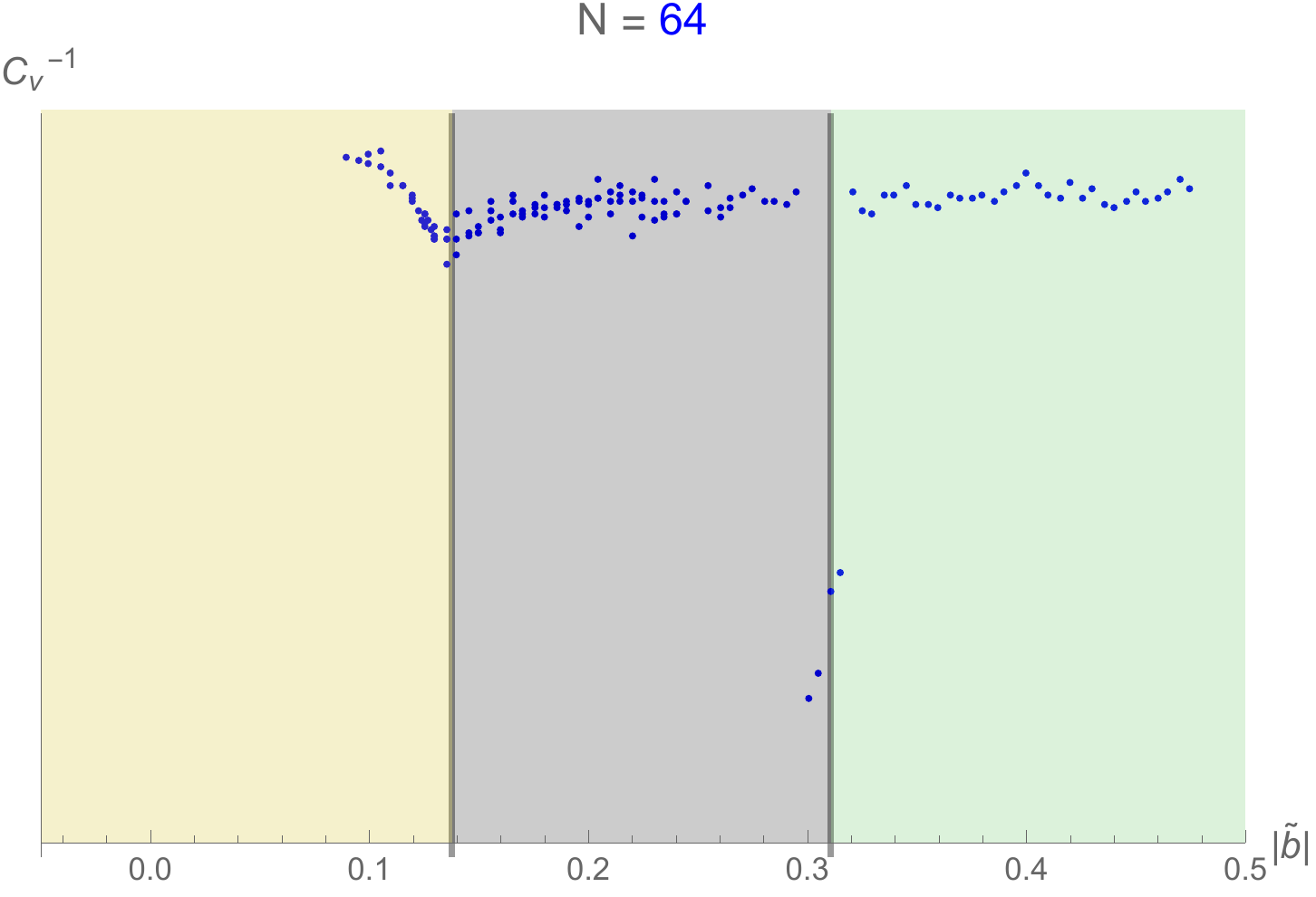} \\

    \includegraphics[width=1\textwidth]{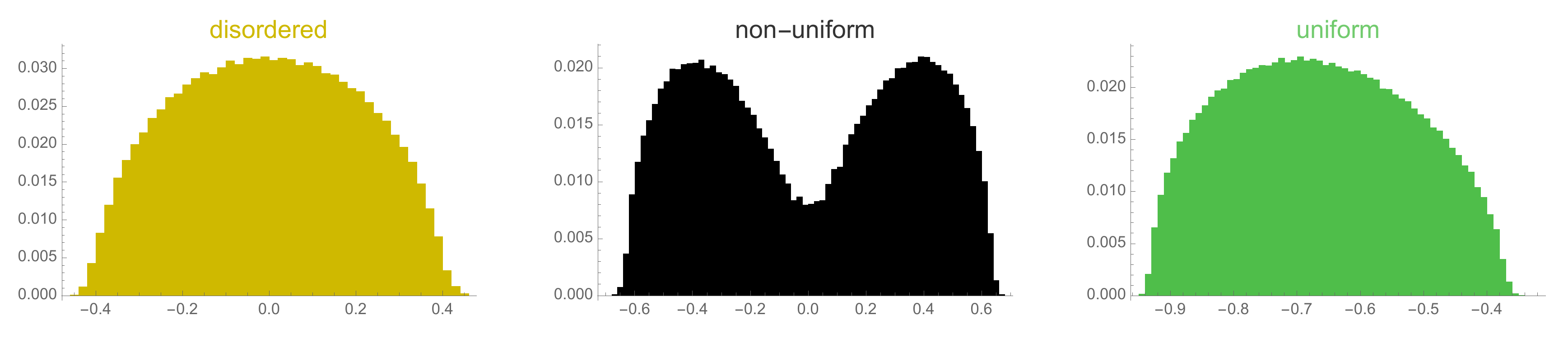}  

  \end{tabular}

  \caption{The top figure shows $1/C_{v}$ for $N=64$ and a range of
    $\tb$ that crosses both transitions. Below are shown typical
    eigenvalue distributions for the respective phases above, where
    the colour coding is matched in the two figures. The distributions
    were taken for $\tc=0.05$ and $-\tb = 0.033333,\, 0.16667$ and $0.4$ respectively
    using the last $4000$ configurations.}
\label{EigenvalueaAndInvCv}
\end{figure}

\section{Conclusion}

We have refined the position of the triple point of a scalar field
theory on a fuzzy sphere by using hybrid Monte Carlo simulations and
extrapolating to the $N \rightarrow \infty$ limit using matrices of
size up to $N=127$. 

We have used three different methods to estimate the position of the triple point, where their confidence intervals overlap we have
\begin{equation}
\tc_T = 0.020 \pm 0.001, \ \tb_T = -0.054 \pm 0.003,
\end{equation}
which is our best estimate for its location.

Our results are in a broad agreement
with recent analytic studies of the model in \cite{Tekel:2017nzf}. It
is conceivable that the true triple point lies at the origin,
however, our results indicate that this is not the case as the origin is
many standard deviations from our estimate.

We have located the phase transitions from peaks in the specific heat
$C_v$, but we also independently checked the eigenvalue distribution
for a range of $\tc$ and various values of $\tb$ to verify the
identified phases. Figure \ref{EigenvalueaAndInvCv} shows the specific
heat for $N=64$ and $\tc = 0.05$ and corresponds to the column of
colour points in Figure \ref{N64PhaseDiag}.

The observed behaviour of the specific heat as the triple point is
approached while taking larger values of $N$, is that the main peak
grows and moves to smaller values of $|\tilde b|$ and a new peak
appears near the large one, but at smaller $|\tb|$, see Figure
\ref{CvSmallN}. These peaks follow the coexistence lines $l_1$ and
$l_3$ of Figure \ref{N64PhaseDiag} and are what we consider the phase
boundaries of the system. While constructing the phase diagram we
have observed similar behaviour further away from the triple
point. More specifically, as we move away from the triple point the secondary
peak gets large and yet another tertiary peak appears, which in turn
grows and bifurcates. This behaviour suggests the existence of
additional triple points further out along the line $l_3$.  We
conjecture that the non-uniformly ordered phase consists of a family
of distinct phases separated by new coexistence lines.  This issue
would deserve a study on its own and we hope to return to it in the
future.

We have not focused on the exact character of the
  phase transitions (in the large $N$ limit), but our simulations show
  clearly that the transition from the non-uniformly ordered to the
  disordered phase is first-order. This can also be seen from the
  rapid growth of the peaks in the specific heat in Figure
  \ref{CvSmallN}. The transition from the disordered to the
  non-uniformly ordered phase appears to be third-order (as in the
  pure potential model) and the transition curve asymptotes to that of
  the matrix model for large $\tilde{c}$. The nature of the transition
  at small $\tilde{c}$ (the disordered to uniform order transition) is
  more difficult to establish and the numerical results are
  inconclusive.  The approximate models studied in
  \cite{Tekel:2017nzf} suggests that this
  transition is of second-order but unfortunately our numerical
  results in this region are not precise enough to support this
  conclusion, but also are not in conflict with it.

Our results are in accord with earlier studies
\cite{GarciaFlores:2009hf} and the approximate models
\cite{Tekel:2017nzf} whose results agree with those shown in Figure
\ref{EigenvalueaAndInvCv}.

The fact that the triple point lies so close to the origin suggests
that one should be able to get reasonable estimates of its location
from direct perturbation theory around zero coupling. We believe that
such an approach is well worth further effort.

A further direction worthy of future study is the introduction of a
higher derivative terms to the gradient term, \cite{Dolan:2001gn}.
These will further regulate the theory and when tuned appropriately,
should push the triple point towards infinity -- leaving only two
phases that should again be in the Ising universality class.

\subsection*{Acknowledgment}
The authors wish to acknowledge the Irish Centre for High-End
Computing (ICHEC) for the provision of computational facilities and
support (Project Name dsphy009c and dsphy010c). The support from
Action MP1405 QSPACE of the COST foundation is gratefully
acknowledged.  S. Kov\'a\v{c}ik was supported by Irish Research
Council funding. D. O'Connor thanks M. Vachovski and S. Kov\'a\v{c}ik thanks to Juraj Tekel for helpful
discussions.

\bibliographystyle{abbrv}

\end{document}